\documentclass{acm_proc_article-sp}

\usepackage[utf8]{inputenc}
\usepackage{color}
\usepackage{xspace}
\usepackage{booktabs}
\usepackage{url}
\usepackage{multicol}
\usepackage{multirow}
\usepackage{subfigure}
\newcommand{\decline}{DecLiNe\xspace}

\newcommand{\TODO}[1]{}		
\newcommand{\TODOTEXT}[1]{}		
\newcommand{\mt}[1]{}
\newcommand{\jk}[1]{}
\newcommand{\jp}[1]{}

\begin{document}

\title{{\ttlit \decline}~-- Models for Decay of Links in Networks}
\numberofauthors{1} 
\author{
  \alignauthor
  Julia Preusse, Jérôme Kunegis, Matthias Thimm, Sergej Sizov \\
  \affaddr{Institute for Web Science and Technologies}\\
  \affaddr{University of Koblenz-Landau, Germany} \\
  \email{\{jpreusse, kunegis, thimm, sizov\}@uni-koblenz.de}
}
\date{7 August 2012}

\maketitle
\TODOTEXT{add hypotheses to approach section\\add missing references\\precision-figure}
\begin{abstract} 
The prediction of graph evolution is an important and challenging
problem in the analysis of networks and of the Web in particular.  But
while the appearance of new links is part of virtually every model of
Web growth, the disappearance of links has received much less attention
in the literature.  To fill this gap, our approach \emph{\decline} (an
acronym for \underline{Dec}ay of \underline{Li}nks in
\underline{Ne}tworks) aims to predict link decay in networks, based on
structural analysis of corresponding graph models.  In analogy to the
link prediction problem, we show that analysis of graph structures can
help to identify indicators for superfluous links under consideration of
common network models. In doing so, we introduce novel metrics that
denote the likelihood of certain links in social graphs to remain in the
network, and combine them with state-of-the-art machine learning methods
for predicting link decay.  Our methods are independent of the
underlying network type, and can be applied to such diverse networks as
the Web, social networks and any other structure representable as a
network, and can be easily combined with case-specific content analysis
and adopted for a variety of social network mining, filtering and
recommendation applications.  In systematic evaluations with large-scale
datasets of Wikipedia we show the practical feasibility of the proposed
structure-based link decay prediction algorithms.
\end{abstract}

\category{H.4}{Information Systems Applications}{Miscellaneous}
\keywords{Web graph, network evolution, link prediction, decay} 

\section{Introduction}
Analysis of link structures is traditionally an important component of Web
information systems, such as search engines, recommender systems, spam
filters, content summarization tools, and many others. These applications
are supported by a wide range of state of the art methods for link-based
authority ranking, prediction of further network evolution, and detection
of structural anomalies. Well-known properties of networks such as the Web
are (1) highly imbalanced distributions of node degrees (in a broader
sense of several existing models, node ``authoritativeness''), and (2)
high clustering coefficient, indicative for existence of multiple or
tightly connected sub-components (``cliques'') \cite{citeulike:1022205}. 
Among many possible use
cases, this knowledge can be used for predicting/suggesting new graph
edges that appear ``reasonable'' in an existing graph structure, e.\,g.,
by connecting two nodes that have many neighbors in common. The prediction
of such ``missing links'' (e.\,g., references between Web pages,
friendships in social networks, followers and citations on
Twitter, cross-references between articles in Wikipedia, etc.) can be
seen as an established recommendation scenario that has been intensively
discussed over the last decade (cf.~\cite{b676}).

In real life, the dynamics of network evolution is more complex and
includes both adding and {\em decay} or removal of connections and relationships.
Unlike link prediction, the issue of link removal has not yet been considered
as a general and domain-independent problem of network
analysis. Our approach coined \decline (an acronym for {\underline Dec}ay of {\underline
Li}nks in online {\underline Ne}tworks) aims to close this gap and to introduce
generic methods for prediction of ``superfluous links'' in networks, based
on structural analysis of corresponding graph models.
\begin{figure}[hbt!]
\centering\epsfig{file=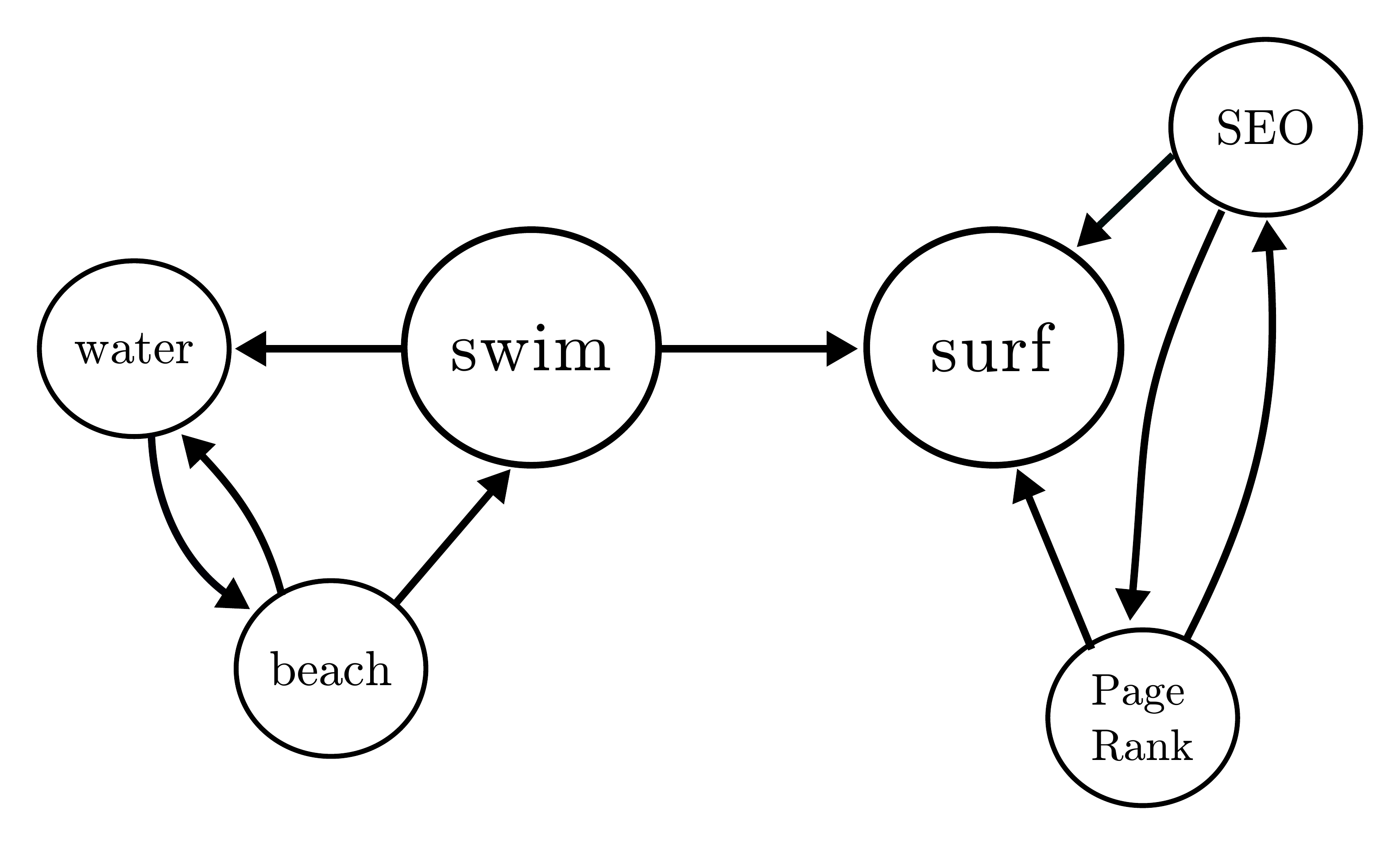, scale=0.18}
\caption{Sample graph $G$ of interlinked Wikipedia articles. The connection between
articles `swim' and `surf' is intuitively wrong.}
\label{fig:swim-surf}
\end{figure}
As a running example, we may consider the fictional graph of a sample of Wikipedia articles from Figure \ref{fig:swim-surf}
\begin{align*}
 G = \{\textit{water}, \textit{swim}, \textit{beach}, \textit{surf}, \textit{SEO}, \textit{PageRank}\}.
 \end{align*}
 A link $(i,j)$ indicates that article $i$ links to article $j$.

The graph $G$ contains two tightly connected components 
\begin{align*}
T_1 &= \{\textit{swim}, \textit{water}, \textit{beach}\} ,\\
T_2 &= \{\textit{surf}, \textit{SEO}, \textit{PageRank}\}.
\end{align*}
The link $(\textit{swim},\textit{surf})$ does not directly belong to structures of $T_1$ and $T_2$
and thus does not connect closely related resources, this can be
recognized by the fact that $(\textit{swim},\textit{surf})$ does not
substantially contribute to the high clustering coefficient of $G$. 
Consequently, we may
assume that the link $(\textit{swim},\textit{surf})$ may demand critical
reconsideration as a potential mistake and will be possibly removed in
the future. 

Conceptually, we discuss in our paper the hypothesis that knowledge of the structure of 
social networks and models allows for defining invariant indicators
for ``superfluous'' links. More precisely, we consider different ways to
solve the link decay problem as a special case of link prediction, by
introducing novel graph models and edge weighting metrics, customized for
prediction of low-likelihood edges.

In our sample graph introduced before, the wrong link has been set due
to missing disambiguation of two meanings for `surf'. In general, the
decision to withdraw a link may have many different reasons and cannot be
fully explained without domain-dependent knowledge about the particular
network and without content resp.\ context analysis of affected nodes (users, web
pages, postings). Our contribution aims to answer the fundamental question:
to what extent can structural analysis contribute to the prediction of
link decay, as a dedicated source of information? The resulting
domain-independent approach of \decline can be easily combined with
case-specific content analysis and adopted for a variety of
applications, such as advanced authority ranking, detection of link spam
and manipulations, recommendations for re-organization of social graphs by
users and content providers, and many others.

The rest of this paper is organized as follows. In Section 2 we discuss
related work in the fields of graph analysis and link prediction. Section
3 formalizes the problem of link decay and introduces novel decay
indicators and prediction methods of \decline. In Section 4 we present
results of systematic evaluations for predictive performance of \decline
on large-scale real data (Wikipedia datasets in several languages).
Section 5 summarizes lessons learned and shows directions of future
research.
\jk{
  Suggestion:  use $\mathbf A$ instead of $A$ for the adjacency matrix. 
}

\section{Related Work}
The problem of recognizing and predicting the decay of links in networks
is related to other problems in the area of network analysis, which we
describe in the following. 

\begin{table}
  \caption{
    Related work about the removal of relationships, classified
    by scenario, and by the features used. 
  }
  \label{tab:rel}
  \centering
  \scalebox{0.97}{
  \begin{tabular}{@{}l|l@{\hspace{-0.4cm}}c@{\;\,}c@{\hspace{-0.15cm}}c@{}}
    \toprule
    \textbf{Scenario} & \textbf{Method} & \multicolumn{3}{c}{\textbf{Features}} \\
    \midrule
    &&  \textbf{Content} & \scalebox{0.92}[1]{\textbf{Interaction}} &  \textbf{Links} \\
    \midrule
    Network & Degrees & -- & -- & \cite{hansAkkermans} \\
    \, evolution & Sequential & -- & \cite{eppstein2002,kleinberg1999} & -- \\
    \midrule
    Decay of &
    Unfriend & \cite{loosingFriends,pew-unfriend} & -- & \cite{loosingFriends} \\
    \, social links & Unfollow & \cite{twitterUnlink} & \cite{twitterUnlink} & \cite{fks2011,twitterUnlink}  \\
    \midrule
    Declining & Groups & -- & \cite{kairam2012} & -- \\
    \, participation & Churn  & -- & \cite{karnstedt:churn} & --  \\
    \midrule
    \multirow{3}{*}{\begin{tabular}{@{}l}Anomaly \\ \, detection\end{tabular}} & Spurious links & -- & -- & \cite{guimera2009,zeng2012}  \\
    & Spam  & \cite{benczur2005} & -- & \cite{benczur2005} \\
    & Disconnection  & -- & \cite{DeRosa:200} & -- \\
    \midrule
    Decay of & Reverts & \cite{adler10,rzeszotarski2012} & \cite{adler10} & -- \\
    \, Web links & Link removal & -- & -- & \textbf{\decline} \\
    \bottomrule
  \end{tabular}
  }
\end{table}

The evolution of networks such as the Web is subject to many models such
as preferential attachment \cite{b439} or the spectral evolution model
\cite{kunegis:spectral-network-evolution}, most of which only model the
addition of edges over time. 
The evolution of the Web hyperlink graph in particular has been studied
too, for instance in \cite{bordino2009}. 
The evolution of the Wikipedia hyperlink graph has been
studied in 2006 \cite{buriol2006}. 
The concrete problem consisting of predicting the appearance of new
links in networks is called \emph{link prediction}
\cite{ASI:ASI20591}.   
Unlike link prediction, the prediction of link disappearance has been
investigated only very little.

\textbf{Network evolution.}
Several graph growth models include link disappearance in addition to
link creation, for instance in a model to explain power laws
\cite{hansAkkermans}.  Other examples can be found in
\cite{eppstein2002} and \cite{kleinberg1999}, in which a model for 
growth of the Web is given in which edges are removed before others are
added. While these methods succeed in predicting global characteristics
of networks such as the degree distribution, they do not model the
structure of the network, and thus cannot be used for predicting
individual links. 

\textbf{Decay of social links.}
For social networks, most studies focus on non-structural reasons for the
disappearance of links, such as interactions between people. 
Examples are the removal of friendships on Facebook (``unfriending'')
\cite{loosingFriends}, and the removal of follow links on Twitter (``unfollowing'')
\cite{kwak2011,twitterUnlink}.
A recent study \cite{pew-unfriend} finds that the most common reason for
unfriending on Facebook is over political opinions.
In all these works, the only structural indicators used for predicting
the disappearance of social links are the number of common neighbors in
\cite{fks2011}, \cite{twitterUnlink} and \cite{loosingFriends}.  All
three studies find that links connecting nodes with many common
neighbors are less likely to be deleted from the social graph, and that
content and interaction features are more predictive for link disappearance in social
networks. Since the form of content and in particular interaction is
fundamentally different among people than hyperlinks between pages, these
methods cannot be generalized to predict the disappearance of
hyperlinks. 

\textbf{Declining participation.}
The decay of groups in social networks is studied in \cite{kairam2012},
explaining it by interaction patterns. 
Another related phenomenon is called \emph{churn}, describing the
situation in which a user quits a social community.  Churn can be
modeled as the deletion of an edge between the user and the service, and
thus corresponds to the deletion of edges in a bipartite graph
\cite{karnstedt:churn}.  
The problems of predicting churn and declining participation are
fundamentally bipartite, since they act on the network connecting users
with items, and are therefore not suited to solving our problem at
hand. 

\textbf{Anomaly detection.}
A related problem is the identification of spurious links, i.\,e., links
that have been erroneously observed~\cite{guimera2009,zeng2012}. 
A related area of research is the detection of link spam on the Web, in which
\emph{bad} links are to be detected \cite{benczur2005}.  
Similarly, the disconnection of nodes has been predicted in mobile ad-hoc networks~\cite{DeRosa:200}. 
These problems
are structurally similar to the problem studied in this paper, but do
not use features that are typical for link prediction such as the
degree of nodes or the number of common neighbors. 

\textbf{Decay of Web links.}
We are not aware of any previous work on the problem of predicting the
disappearance of links on the Web. 
In the context of Wikipedia, 
a problem related to ours is the identification of reverts.  These have
been predicted in previous work 
by giving each word a score based on how likely it is to be
reverted \cite{rzeszotarski2012}, or alternatively by measuring the quality
of edits \cite{adler10}.  However, none of these related works considered the
reverting of wikilinks. 

An overview of the methods is given in Table~\ref{tab:rel}.  We show the
methods according to three types of features that are used: (a)~Content
information, e.\,g., when the content of Facebook posts is used as an
indicator for unfriending, (b)~Interaction information, e.\,g., when the
decline of wall-writing on Facebook is used as an indicator for
unfriending, and (c)~Link information, e.\,g., when a low number of common
friends is used as an indicator of a likely unfriending. 
In summary, we can state that the \decline approach is the first approach using
link information for explaining the disappearance of links on the
Web. 

\section{The \decline approach}
In the following, we investigate the problem of predicting decay of links in networks in a general and formal manner. 
Depending on the type of a network, removal of links may be caused by different issues. 
In general, the reasons for a link being removed may be content-based reasons, e.\,g., a hyperlink from a Wikipedia page is removed as the articles' 
topics are not related, 
structural reasons, e.\,g., removing a network link in a telecommunications network, or a combination of both. 
For our treatment we consider only structural properties of the underlying network and we do this for two reasons. 
First, our objective is to find general domain-independent models, whereas content is clearly domain-dependent.
Second, we hypothesize that several content-based reasons are also reflected in the network structure. 
Coming back to our introductory example from Figure \ref{fig:swim-surf}, two different main topics can be found and are manifested in the two highly-connected components and only one link between.
Although this hypothesis is, of course, not generally applicable we focus on structural properties in order to investigate how good we can predict decay of links without considering content.
\subsection{Problem Formalization}
A network $\mathbf{N}$ is defined as
\begin{align*}
\mathbf{N} = (V,E),
\end{align*}
where $V$ is a set of nodes or items and $E$ is a set of edges, $E \subseteq V \times V$. 
In order to predict decay of links we consider a scenario of evolving networks. 
Let 
\begin{align*}
\mathbf{N}_t = (V_t,E_t)
\end{align*} 
for $t\in\mathbb{N}$ be the network $\mathbf{N}_{t}$ at time $t$ with $V_{t}$ being the set of nodes of $\mathbf{N}_{t}$ and $E_{t}\subseteq V_{t}\times V_{t}$ the set of links of $\mathbf{N}_{t}$. 
Without loss of generality, we assume that $V_{t}=V_{t'}$ for all $t,t'\in\mathbb{N}$, otherwise we could define $V_{1}\cup V_{2}\ldots$ to be the set of nodes for each network. 
We also write $\mathbf{N}_t = (V,E_t)$ for $t\in\mathbb{N}$ and define $n=|V|$. 

Typically, a network $\mathbf{N}_{t}$ is represented by its adjacency matrix $\mathbf{A}(\mathbf{N}_t)$, i.\,e., $V$ is defined via $V=\{1,\ldots,n\}$ and $\mathbf{A}(\mathbf{N}_{t}) \in \{0,1\}^{n \times n}$ is defined as
\begin{align*}
\mathbf{A}(\mathbf{N}_t)_{ij} = \begin{cases}
1, \quad \mbox{if } (i,j) \in E_t,\\
0, \quad \mbox{otherwise}.
\end{cases}
\end{align*}
If the actual network and evolution step is of no importance we usually write $\mathbf{A}$ instead of $\mathbf{A}(\mathbf{N}_{t})$.

\textbf{Link prediction.}
A link prediction function $f_m$ is a function
\begin{align}
  f_m: \{0,1\}^{n \times n} \rightarrow \mathbb{R}^{n \times n}
  \label{eq:link}
\end{align}
that takes a matrix $\mathbf{A}(\mathbf{N}_t)$ and assigns for each node pair $i,j \in V$ a link creation score by computing measure $m$ \cite{ASI:ASI20591}.
The bigger a link prediction score of an edge $(i,j) \notin E_t$ is, the more it is expected to actually be added to the network.
Thus, good link prediction functions assign larger scores to links $(i,j)$ that will appear until time $t+1$, i.e. $(i,j) \in E_{t+1} \setminus E_t$, than to others.

For the problem of predicting link decay, our aim is to define a link decay score function $g_m$ of the form
\begin{align}
  g_s: \{0,1\}^{n \times n} \rightarrow \mathbb{R}^{n \times n}
  \label{def:unlink}
\end{align}
that takes a matrix $\mathbf{A}(\mathbf{N}_{t})$ and computes for each node pair a decay score by measure $m$. 
More specifically, for edges $(i,j) \in E_{t}\setminus E_{t+1}$ we expect $g_m(\mathbf{N}(\mathbf{A}_{t}))_{ij}$ to be significantly larger than other decay scores of other edges.
\subsection{Predictive Models}
The problem of predicting whether a link decays can be viewed as the inverse problem of predicting the creation of links, which is also known as the link prediction problem. 
The objective of \decline is to validate how far link decay can be predicted with the same structural methods as link prediction.
In the following, we propose two different approaches for answering this question. 
These approaches complement link prediction by complementing the score (cf.\ Section \ref{sec:model1}) and the network (cf.\ Section \ref{sec:model2}), respectively.
\begin{table*}[hbt!]

  \centering
  \caption{
    Overview of all score methods for link and link decay prediction of an edge $(i,j)$
  }
  \begin{tabular}{ l @{\qquad} r@{$=$}l @{\qquad} l @{\,} l @{\qquad} l @{\,} l }
    \toprule
    \textbf{Name} & \multicolumn{2}{l}{\textbf{Link prediction function}} & \multicolumn{2}{l}{\textbf{Inverse}} & \multicolumn{2}{l}{\textbf{Complement}}\\
    \midrule
    Preferential attachment & $f_{\textit{PA}}(\mathbf{A})_{ij} $&$\delta(i) \cdot \delta(j)$ & $-f_{\textit{PA}}(\mathbf{A})_{ij}$ & [Eq.~\eqref{eq:PAU1}]& $f_{\textit{PA}}(\bar{\mathbf{A}})_{ij}$& [Eq.~\eqref{eq:PA}] \\
    Common neighbors & $f_{\textit{CN}}(\mathbf{A})_{ij} $&$ (\mathbf{A}^2)_{ij}$ & $-f_{\textit{CN}}(\mathbf{A})_{ij}$ & [Eq.~\eqref{eq:CNU1}] & $f_{\textit{CN}}(\bar{\mathbf{A}})_{ij}$ & [Eq.~\eqref{eq:CN}] \\
    Cosine similarity & $f_{\textit{cos}}(\mathbf{A})_{ij} $&$ \frac{(\mathbf{A}^2)_{ij}}{\sqrt{\delta(i)} \cdot \sqrt{\delta(j)}}$ & $-f_{\textit{cos}}(\mathbf{A})_{ij}$ & [Eq.~\eqref{eq:cosU1}] & $f_{\textit{cos}}(\bar{\mathbf{A}})_{ij}$& [Eq.~\eqref{eq:cos}] \\
    Jaccard index & $f_{\textit{Jacc}}(\mathbf{A})_{ij} $&$ \frac{(\mathbf{A}^2)_{ij}}{|N(i)\cup N(j)|}$ &  $-f_{\textit{Jacc}}(\mathbf{A})_{ij}$ & [Eq.~\eqref{eq:JaccU1}] & $f_{\textit{Jacc}}(\bar{\mathbf{A}})_{ij}$& [Eq.~\eqref{eq:Jacc}] \\
    Adamic--Adar & $f_{\textit{Adad}}(\mathbf{A})_{ij} $&$ \sum_{k \in
      N(i) \cap  N(j)}\frac{1}{\log \delta(k)}$ &  $-f_{\textit{Adad}}(\mathbf{A})_{ij}$ & [Eq.~\eqref{eq:AdadU1}] & $f_{\textit{Adad}}(\bar{\mathbf{A}})_{ij}$& [Eq.~\eqref{eq:Adad}] \\
    \bottomrule
  \end{tabular}
  \label{table:scores}
\end{table*}

\subsubsection{Model 1: Complement Score}
\label{sec:model1}
 Using a link prediction function $f_m$ from \eqref{eq:link} that computes a score by measure $m$ we define its \emph{inverse link prediction function} $g_m^1$ via
\begin{align*}
	g^1_m(\mathbf{A}) & = - f_{m}(\mathbf{A})\qquad.
\end{align*}
The rationale behind this complement model is that links that have a high link prediction score should not be removed, whereas links with a low score are expected to be deleted.
In the literature a series of different approaches have been proposed for solving the link prediction problem \cite{b676}. In this paper we consider the following approaches as the basis for unlink prediction.
\paragraph{Preferential attachment} 
Let $\delta(i)$ denote the degree of node $i$ and let $ \delta(j)$ denote the degree of a node $j$ in $\mathbf{A}$. Preferential attachment estimates that an edge $(i,j)$ is added with a likelihood proportional to the product of the degree of $i$ and the degree of $j$, i.e., we have $f_{\textit{PA}}(\mathbf{A})_{ij} = \delta(i) \cdot \delta(j)$.
Hence, the \textit{complement score} score of $(i,j)$ is
\begin{align}
g^1_{\textit{PA}}(\mathbf{A})_{ij} &= - \delta(i) \cdot \delta(j).
\label{eq:PAU1}
\end{align}
Thus according to this method, links are likelier to be removed between two nodes of a low degree.
\paragraph{Common neighbors} 
This link predication method implements the intuition that two nodes are to be linked if they share a lot of neighbors. The function $f_{\textit{CN}}$ is defined via $f_{\textit{CN}}(\mathbf{A})_{ij}= (\mathbf{A}^2)_{ij}$, where $(\mathbf{A}^2)_{ij}$ is the number of paths of length 2 between $i$ and $j$, i.\,e., the common neighbors.
$g^1_{\textit{CN}}$ is therefore defined as
\begin{align}
g^1_{\textit{CN}}(\mathbf{A})_{ij} &= - (\mathbf{A}^2)_{ij}
\label{eq:CNU1}
\end{align}
Links in this model are expected to be removed if they have only few common neighbors.
\paragraph{Cosine similarity} 
With the cosine similarity method, an edge $(i,j)$ is estimated to be created with likelihood proportional to the angle between the degree vectors of node $i$ and $j$.
$f_{\textit{cos}}$ and $g^1_{\textit{cos}}$ are defined as
\begin{align}
	 g^1_{\textit{cos}}(\mathbf{A})_{ij} = - f_{\textit{cos}}(\mathbf{A})_{ij} &= - \frac{(\mathbf{A}^2)_{ij}}{\sqrt{\delta(i)} \cdot \sqrt{\delta(j)}}.
	 \label{eq:cosU1}
\end{align}
If the two nodes are connected to the same nodes, the link between them is expected to stay.

\paragraph{Jaccard index} 
Let $N(k)$ be the set of \emph{neighbors} of node $k \in V$, i.\,e.,
\begin{align*}
	N(k) & =\{l\in V\mid \mathbf{A}_{kl} = 1\}
\end{align*}
With the Jaccard index, an edge is created with likelihood proportional
to the number of common neighbors divided by the number of different
neighbors of both nodes. The function $f_{\textit{Jacc}}$ and the
corresponding function $g^1_{\textit{Jacc}}$ are defined via 
\begin{align}
    g^1_{\textit{Jacc}}(\mathbf{A})_{ij} = f_{\textit{Jacc}}(\mathbf{A})_{ij} &= - \frac{(\mathbf{A}^2)_{ij}}{|N^{\mathbf{A}}(i)\cup N^{\mathbf{A}}(j)|}.
  \label{eq:JaccU1}
\end{align}
If two nodes are not connected to many nodes but share only few common nodes, the link between them is expected to decay.

\paragraph{Adamic--Adar} 
The measure used by the approach of Adamic and Adar \cite{b475} counts
the number of neighbors of nodes $i$ and $j$, weighted by the inverse
logarithm of each neighbor $k$'s degree $\delta(k)$:
\begin{align}
  g^1_{\textit{Adad}}(\mathbf{A})_{ij} =
  f_{\textit{Adad}}(\mathbf{A})_{ij} &= - \sum_{k \in N(i) \cap
    N(j)}\frac{1}{\log \delta(k)}.
  \label{eq:AdadU1}
\end{align}
Thus, if two nodes share only few common neighbors with a high degree, the link between them is not expected to stay in the network. 


\subsubsection{Model 2: Complement Network}
\label{sec:model2}
\begin{figure}[hbt!]
\centering
\epsfig{file=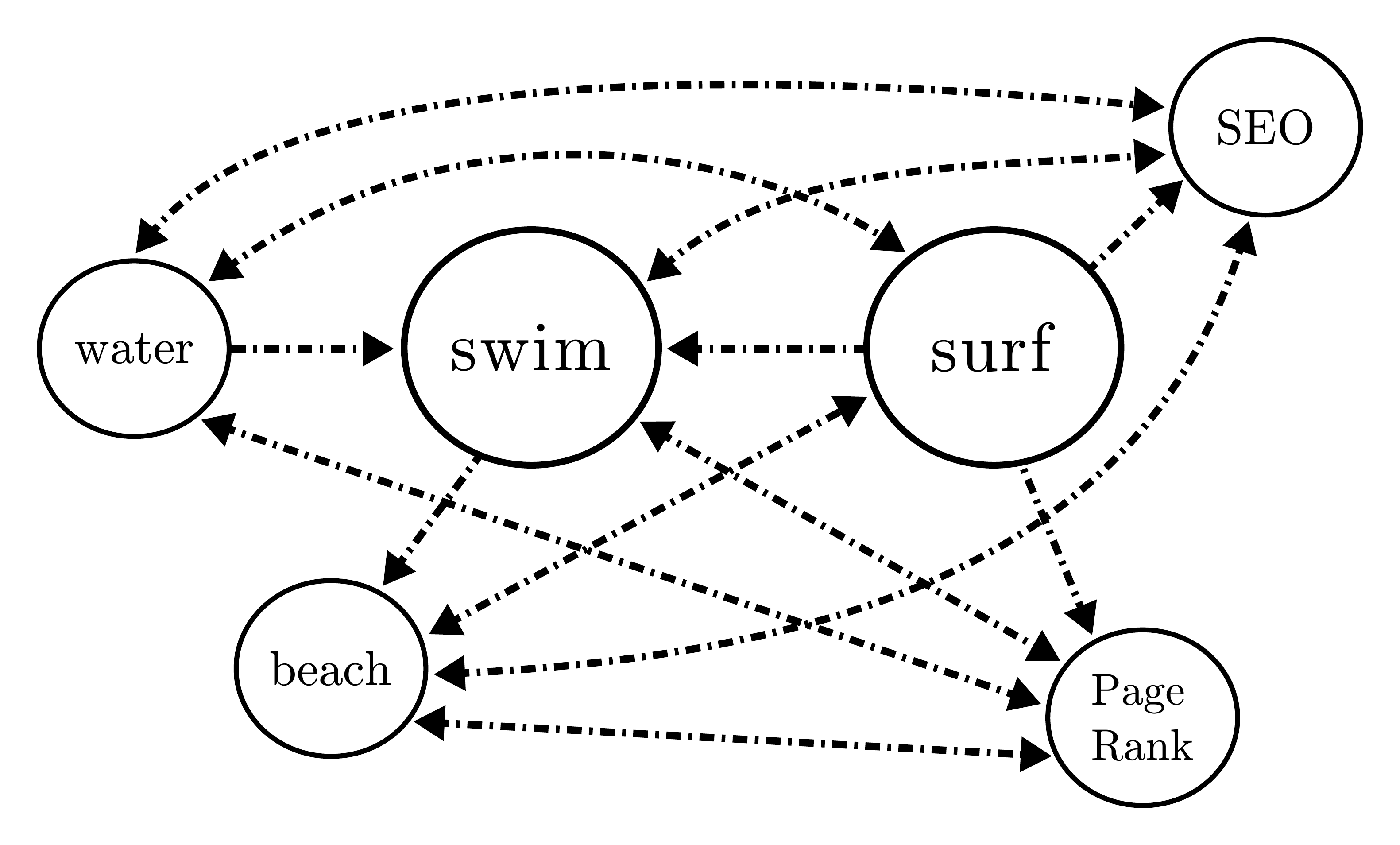, scale=0.18}
\caption{Complement network of network in Figure \ref{fig:swim-surf}. It consists of all edges that are not present in the original network.}
\label{fig:com-swim-surf}
\end{figure}

The second family of link decay functions we consider employs link prediction functions as well. But rather than inverting the prediction function we now invert the problem itself and consider predicting removal of links in a network by predicting creation of links in its \emph{complement network}. 
Using a link prediction function $f_m$ we define its complement link prediction function $g^2_m$ via
\begin{align*}
	g^2_m(\mathbf{A}) & =  f_m(\bar{\mathbf{A}})\qquad.
\end{align*}
Given a network $N=(V,E)$ its complement $\bar{N}=(V,\bar{E})$ is defined via $\bar{E}=\{(i,j)\mid i\neq j, (i,j)\notin E\}$, i.e., $\bar{N}$ contains only links between different nodes that are not connected in $N$.
 The complement network of the network in Figure \ref{fig:swim-surf} is shown in Figure \ref{fig:com-swim-surf}.
The rationale behind this complement model is that since it contains all
non-edges, edges that are predicted in it, should not be present in the
original network. Thus, we can conclude the likelihood with which they
can be removed.  The complement network is by far not sparse, thus we
cannot represent the complement network as a matrix.  Since link prediction
methods compute a score of a network's adjacency matrix, we will use the
following alternative that does not need the adjacency matrix of the
complement graph to be constructed.  
If $\mathbf{A}=\mathbf{A}(N)$ is the adjacency matrix
of $N$ then $\bar{\mathbf{A}}=\mathbf{A}(\bar{N})$ can be written as
\begin{align}
	\bar{\mathbf{A}} & = \mathbf{1}-I-\mathbf{A}
	\label{eq:complement}
\end{align}
where $\mathbf{1}$ is the $1$-matrix (containing only $1$s) and $I$ is the identity matrix (containing $1$s in the diagonal). 

We expect that predicting creation of links in $\bar{\mathbf{A}}$ also solves the problem of predicting removal of links in $\mathbf{A}$. 
Considering Figure \ref{fig:com-swim-surf} again, we can see that predicting a link between nodes `swim' and `surf' is very likely, e.\,g., using $g^2_\textit{CN}$. From the prediction of this edge in the complement network, its decay in the original network would be predicted.

In the following, we use Equation \eqref{eq:complement} to derive $g^2_m(A)_{ij}$ using the same link prediction measures $m$ as in the previous section.

\paragraph{Preferential attachment}
An edge $(i,j)$ is removed with a likelihood proportional to product of the degree of node $i$ and degree of node $j$ in the complement network $\bar{N}$.
\begin{align}
  g^2_{\textit{PA}}({\mathbf{A}})_{ij}& = f_{\textit{PA}}(\bar{\mathbf{A}})_{ij}\notag\\
  &= (n-1- \delta(i)) \cdot (n-1-\delta(j))
  \label{eq:PA}
\end{align}
A link is therefore likely to disappear between low-degree nodes.
\paragraph{Common neighbors}
The link decay score of an edge $(i,j)$ in the original network
is then translated to the link prediction score in its complement network by
\begin{align}
	g^2_{\textit{CN}}({\mathbf{A}})_{ij} &= f_{\textit{CN}}(\bar{\mathbf{A}})_{ij}\notag\\
	&=  n - \delta(i) - \delta(j) + (\mathbf{A}^2)_{ij}.
\label{eq:CN}
\end{align}
Thus, a link is likely to stay if the degrees of its incident nodes are big and share many neighbors.

\paragraph{Cosine similarity}
An edge is removed with a likelihood proportional to the angle between the complemented degree vectors.
\begin{align}
  g^2_{\textit{cos}}({\mathbf{A}})_{ij} &= f_{\textit{cos}}(\bar{\mathbf{A}})_{ij}\notag\\
  &= \frac{n - \delta(i) - \delta(j) +  (\mathbf{A}^2)_{ij} }{\sqrt{(n - 1 - \delta(i))}\cdot \sqrt{(n - 1 - \delta(j))}}
  \label{eq:cos}
\end{align}

\paragraph{Jaccard index}
The Jaccard measure computes the score of edge $(i,j)$ by the ratio of number of common neighbors and numbers of nodes that are adjacent to $i$ or $j$.
Applied to the complement network, we obtain the following link decay score
\begin{align}
  g^2_{\textit{Jacc}}(\mathbf{A})_{ij} &= f_{\textit{Jacc}}(\bar{\mathbf{A}})_{ij}\notag\\
  &= \frac{n - \delta(i) - \delta(j)  + (\mathbf{A}^2)_{ij} }{|N(i) \cup N(j)|}.
  \label{eq:Jacc}
\end{align}
According to this measure, an edge is expected to be removed if the degrees of its incident nodes are small and have more dissimilar neighbors.
\paragraph{Adamic--Adar} 
The weighted variant of the Adamic--Adar score of the complement
network is as follows 
\begin{align}
  g^2_{\textit{Adad}}(\mathbf{A})_{ij} &=
  f_{\textit{Adad}}(\bar{\mathbf{A}})_{ij} \notag\\ 
  &= \sum_{k \in V}\frac{1}{\log \delta(v)} - \sum_{k \in
    N(i)}\frac{1}{\log \delta(k)}
  - \sum_{k \in N(j)}\frac{1}{\log \delta(k)} \notag\\
  &+ \sum_{k \in N(i) \cap N(j)}\frac{1}{\log \delta(k)}.
  \label{eq:Adad}
\end{align}
Under this model, 
if nodes $i$ and $j$ are adjacent to few and rather high-degree nodes
the link $(i,j)$ is likely to decay. 

A summary of the scoring methods is given in Table \ref{table:scores}.

\subsection{Predictions in Directed Networks}
The link prediction and link decay methods in this section were aligned for undirected networks, so they used characteristics such as degree $\delta(i)$ and neighborhood $N(i)$ of a node $i$. 
For \decline we evaluate methods and models for link decay predictions on \emph{directed} Wikipedia article-hyperlink networks.
Instead of only one node degree for undirected networks, three different degrees of a node can be defined for directed networks: a node's out- respectively in-degree and its degree.
\begin{figure}[hbt!]
\centering\epsfig{file=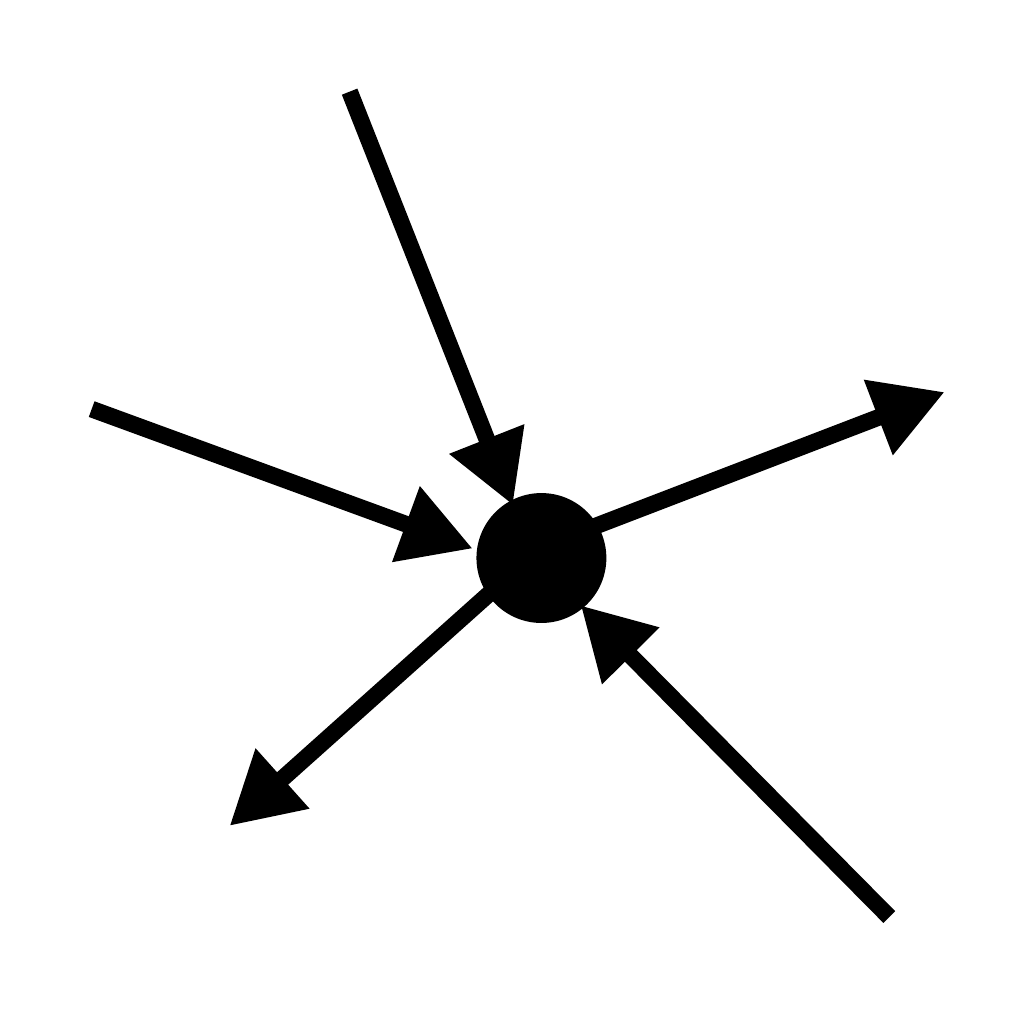, scale=0.18}
\caption{An arbitrary node $i$ with incoming and outgoing edges.}
\label{fig:node}
\end{figure}
Consider the node shown in Figure \ref{fig:node}. It's out-degree $\delta_{\textit{out}}$ is defined as the number of outgoing links from it and its in-degree $\delta_{\textit{in}}$ is defined as the number of incoming links. 
For the given node $i$, $\delta_{\textit{out}}(i) = 2$ and $\delta_{\textit{in}}(i) = 3$. The degree $\delta$ is defined as $\delta_{\textit{out}} + \delta_{\textit{in}}$, so $\delta(i) = 5$.
Further, the node neighborhood $N(i)$  of a node $i$ can now be defined for outgoing and incoming links accordingly
\begin{align*}
N_{\textit{out}}(i) = \{j \in V \mid (i,j) \in E\}\\
N_{\textit{in}}(i) = \{j \in V \mid (i,j) \in E\}.
\end{align*}
A common approach when predicting links in directed network is to use the same methods as for undirected networks but to test different degree combinations \cite{b670}. Thus, all undirected degrees $\delta(i)$ and $\delta(j)$ are aligned with all given combinations from Table \ref{table:degrees}.
\begin{table}

  \centering
  \caption{
    List of the combinations of degrees of node $i$ and node $j$ used.
  }

\begin{tabular}{c | c @{\qquad} c @{\qquad} c @{\qquad} c}
\toprule
\textbf{Name} & \textbf{sym} & \textbf{asym} & \textbf{in} & \textbf{out}\\
\midrule
$\delta_1(i)$ & $\delta(i)$ &  $\delta_{\textit{out}}(i)$ & $\delta_{\textit{in}}(i)$ & $\delta_{\textit{out}}(i)$\\
$\delta_2(i)$ & $\delta(j)$ &  $\delta_{\textit{in}}(j)$ & $\delta_{\textit{in}}(j)$ & $\delta_{\textit{out}}(j)$\\
\bottomrule
\end{tabular}
\label{table:degrees}
\end{table}

For better readability, the methods in this section were aligned with the 'sym' degree (column 1 in Table \ref{table:degrees}) version only. 
Other methods can be defined analogously and have been systematically tested in this work.

\section{Evaluation}
By utilizing common link prediction methods we have defined two families of approaches to predict decay in networks. In this section we conduct an empirical evaluation on how good our approaches work on real datasets. In particular, we stipulate that, given the evolution of some network, links that are removed in a step of the evolution receive a high link decay score. Furthermore, given that we approach the problem of predicting removal of links by using link prediction methods we ask the question of how related those two problems are in real datasets and if they can be solved using the same methods.
We conduct our analysis using five directed large-scale networks from Wikipedia.
As general practice, we evaluate link decay methods for directed networks with different combinations of in-degree and out-degree \cite{b670}.  
Thus, we will explore which effects the different degree combinations have on the prediction quality and which prediction method provides the best precision.
\subsection{Datasets}
To evaluate our proposed decay models, we use the directed
article-hyperlink networks of five of the six
largest\footnote{\url{http://meta.wikimedia.org/wiki/List_of_Wikipedias}}$^{,}$\footnote{The
  evaluation of this dataset is currently ongoing and may be added to a
  later revision of this paper} Wiki\-pedias.  
the English Wikipedia, due to its size and limited computational
resources.  In the directed article-hyperlink network of Wiki\-pedia, a
link between two articles $i$ and $j$ is present if article $i$ links to
article $j$.  
For our link decay prediction scenario we omit user pages and article
discussion pages.

We use the Wikipedia dataset as it resembles the link structure of Web
and is more easy to observe than the latter \cite{buriol2006}. 
The differences between the Wikipedia hyperlink structure and that of the Web is studied in \cite{kamps2009} where Wikipedia is found to be denser and that
outlinks correlate more with page relevance. 

\begin{figure}[hbt!]
  \centering
  \epsfig{file=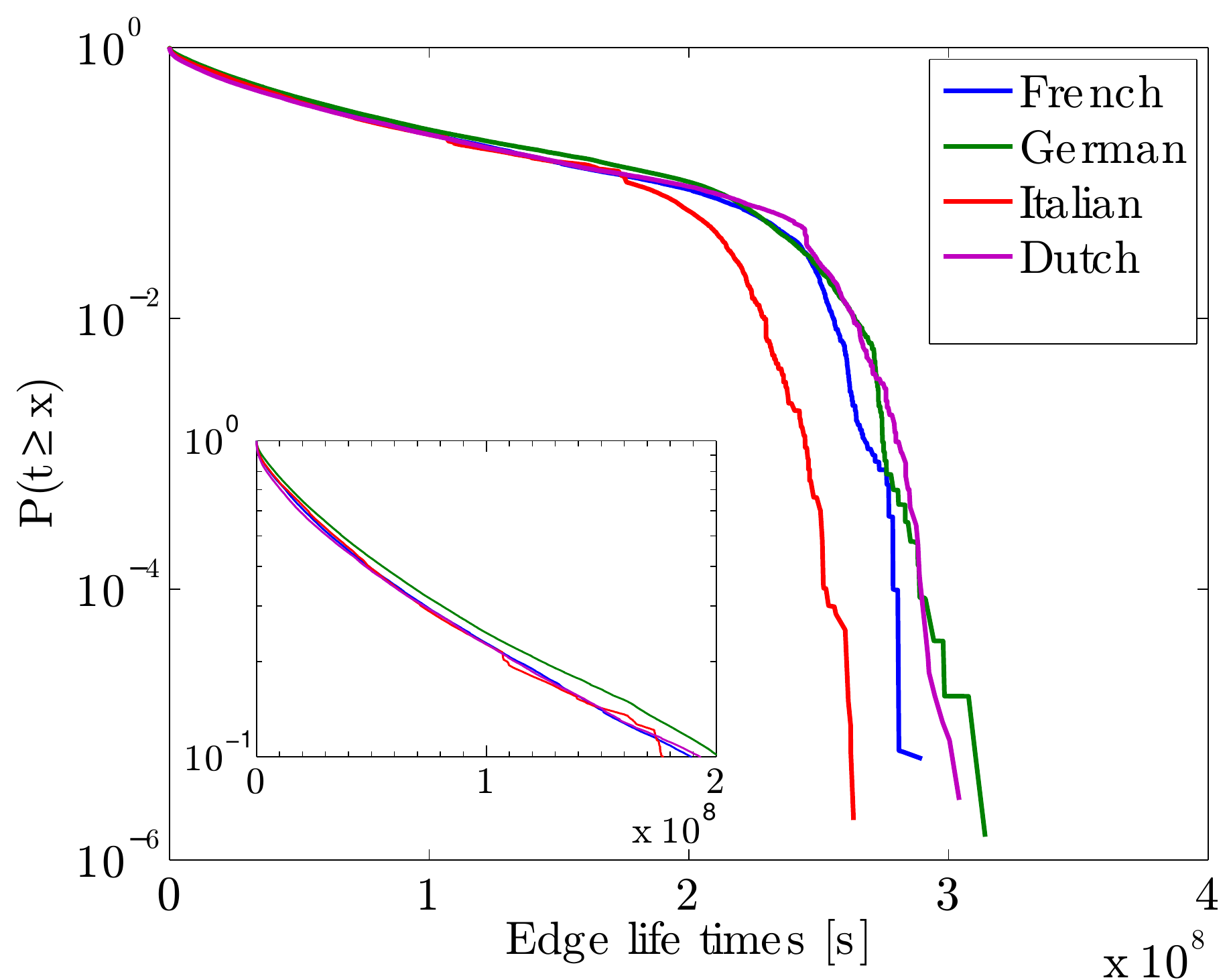, scale=0.4}
  \caption{
    The decay of edges in the five studied Wikipedias is exponential~--
    this means that the probability that an edge will remain for a certain
    time $t$ is proportional to $2^{-t/t_{0.5}}$, where $t_{0.5}$ is the
    half-life of about 23 months. 
  }
  \label{fig:decayDis}
\end{figure}

For each of the five Wikipedias we considered all creation and deletion
events for links since their installment. An over\-view over the
datasets is given in Table \ref{table:datasets}. 
The French Wikipedia is the biggest dataset used with around 1.8 million articles between which overall 41.7 million links where added and 17.3 million removed.
Note that the number of articles includes also articles that where removed later. 
For these Wiki\-pe\-dias, link deletions make up about 24--31\% of all link operations, thus accounting for a large part of structural changes. As shown in Figure~\ref{fig:decayDis}, the decay of links follows an
exponential distribution with a half-life of about 23 months.
The non-exponential behavior of the
decay for $t > 2\times 10^8s$ is due to the young age of Wikipedia;
this time corresponds to about six years, while Wikipedia is 11
years old.  
Notably, for the exponential distribution the expected lifetime of a particular article remains constant over time (memorylessness).
Consequently, the age of an edge cannot be used as an indicator for decay.

\begin{table}

  \centering
  \caption{
    The datasets used in our evaluation. The number of articles also includes articles that were removed.
  }
  \begin{tabular}{lrrr} 
    \toprule
    \textbf{Wikipedia} & \textbf{\#Articles} & \textbf{Adds } & \textbf{Deletes } \\ 
    & & [$\times 10^6$] & [$\times 10^6$] \\
    \midrule
    French & 1,763,659 & 41.7 & 17.3 \\ 
    German & 1,526,219 & 58.7 & 27.6\\ 
    Italian & 953,208 & 26.0 & 8.9 \\ 
    Polish & 765,930 & 18.8 & 6.2\\ 
    Dutch & 751,888 & 15.3 & 4.7\\
    \bottomrule
  \end{tabular}
\label{table:datasets}
\end{table}

\subsection{Methodology}
\begin{figure*}[th!]
\centering
\epsfig{file=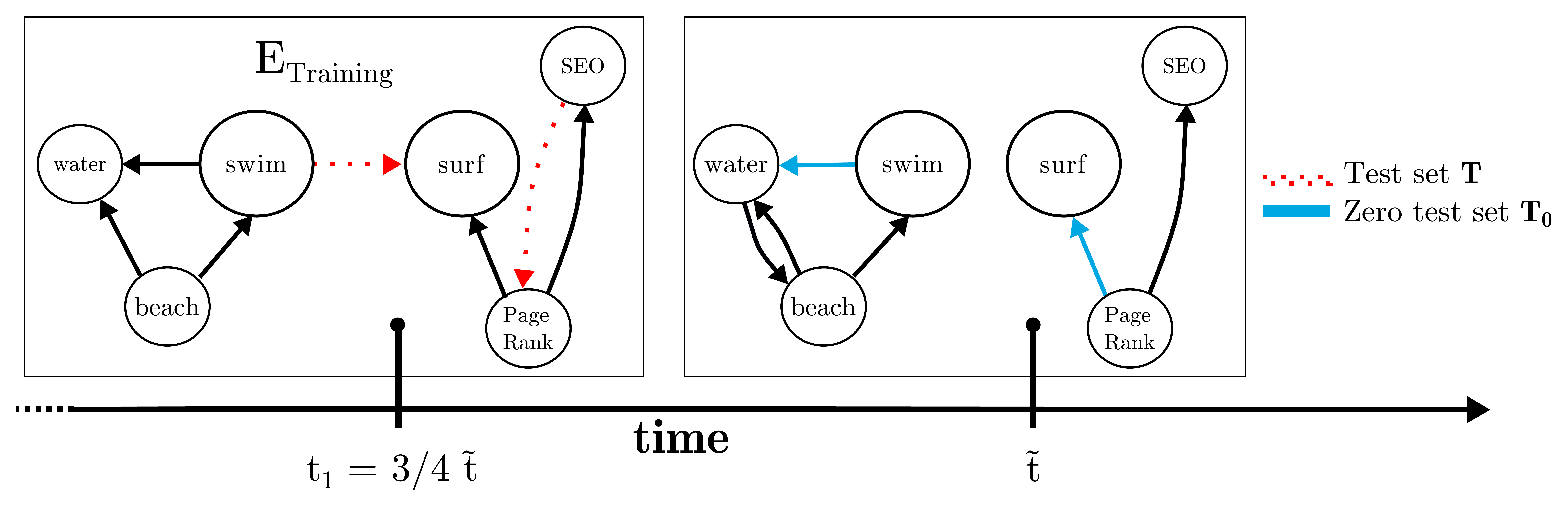, scale=0.15}
\caption{Split in training and test set.}
\label{fig:split_trainingTest}
\end{figure*}

In our evaluation we aim to compare how well we can distinguish edges that have been removed and edges that are not removed.
For that we split the datasets of a Wikipedia article network $\mathbf{N} = (V,E)$ at time point $t_1 = 3/4\tilde{t}$ of the whole time interval $\tilde{t}$. We define the training set as all edges that are present at time point $t_1$
\begin{align*}
 E_{\textit{Training}} = E_{t_1}, 
\end{align*}
the test set $\mathbb{T}$ as all edges from the training set that are not present anymore at time $\tilde{t}$
\begin{align*}
\mathbb{T} = \{(i,j) \in E_{t_1}\setminus E_{\tilde{t}} \mid i,j\in V\}, 
\end{align*}
and the zero test set $\mathbb{T}_0$ as random sample of edges from the training set that are still present at time $\tilde{t}$ with size $|\mathbb{T}_0| = |\mathbb{T}|$
\begin{align*}
\mathbb{T}_0 = \{(i,j) \in E_{t_1}\setminus E_{\tilde{t}} \mid i,j\in V\}.
\end{align*}
The three edge sets are illustrated in Figure \ref{fig:split_trainingTest}.

We compute the precision of our models with the average precision measure, which is defined as follows.
Given edges from test set $\mathbb{T}$ and zero test set $\mathbb{T}_0$ and link decay scores $g_{ij}$ for all edges $(i,j)$ from these two sets, we produce a ranking $R$ of all edges $(i,j)$ by sorting them in descending order. Thus, $R(1)$ is the edge with the highest link decay score and $R(l)$ with $l = |\mathbb{T}| + |\mathbb{T}_0|$ represents the lowest scored edge. 

Then, the average precision \textit{AP} is defined as
\begin{align*}
\textit{AP} &= \frac{\sum_{i = 1}^l{P(i) * I(i)}}{|\mathbb{T}|},
\end{align*}
where $I$ is an indicator function defined as 
\begin{align*}
I = \begin{cases}
1, \quad \text{if~} i \in \mathbb{T}, \\
0, \quad \text{if~} i \in \mathbb{T}_0\\
\end{cases}
\end{align*}
and $P(i)$, the precision at cut-off $i$, is defined as
\begin{align*}
P(i) &= \frac{|\mathbb{T} \cap \{j | R(j) \leq i\} |}{i}.
\end{align*}
By construction, the precision of the random baseline --~which predicts
every edge to be removed with a probability of $0.5$~-- is thus $0.5$.

We compute the average precision for all combinations of link decay scores shown in Table \ref{table:scores} and the four combinations of degrees from Table \ref{table:degrees} for the five largest Wikipedias.
Analysis code as well as the datasets will be made available at \url{konect.uni-koblenz.de/research/decline}.
\subsection{Results}
In the following, we provide results of our empirical evaluations.
\paragraph{Precision of decay models}
In Section 3 we have defined two decay models that transform the link prediction problem to the problem of predicting link removal.
Each of the two decay models computes scores of five classic link prediction methods: \textit{preferential attachment} (PA), \textit{common neighbors} (CN), \textit{cosine} (cos), \textit{Jaccard} (Jacc), and \textit{Adamic--Adar} (Adad), which in turn are varied by four different out and in-degree combinations.
\begin{figure*}[htb!]
\centering
\subfigure[Complement score]{
\epsfig{file=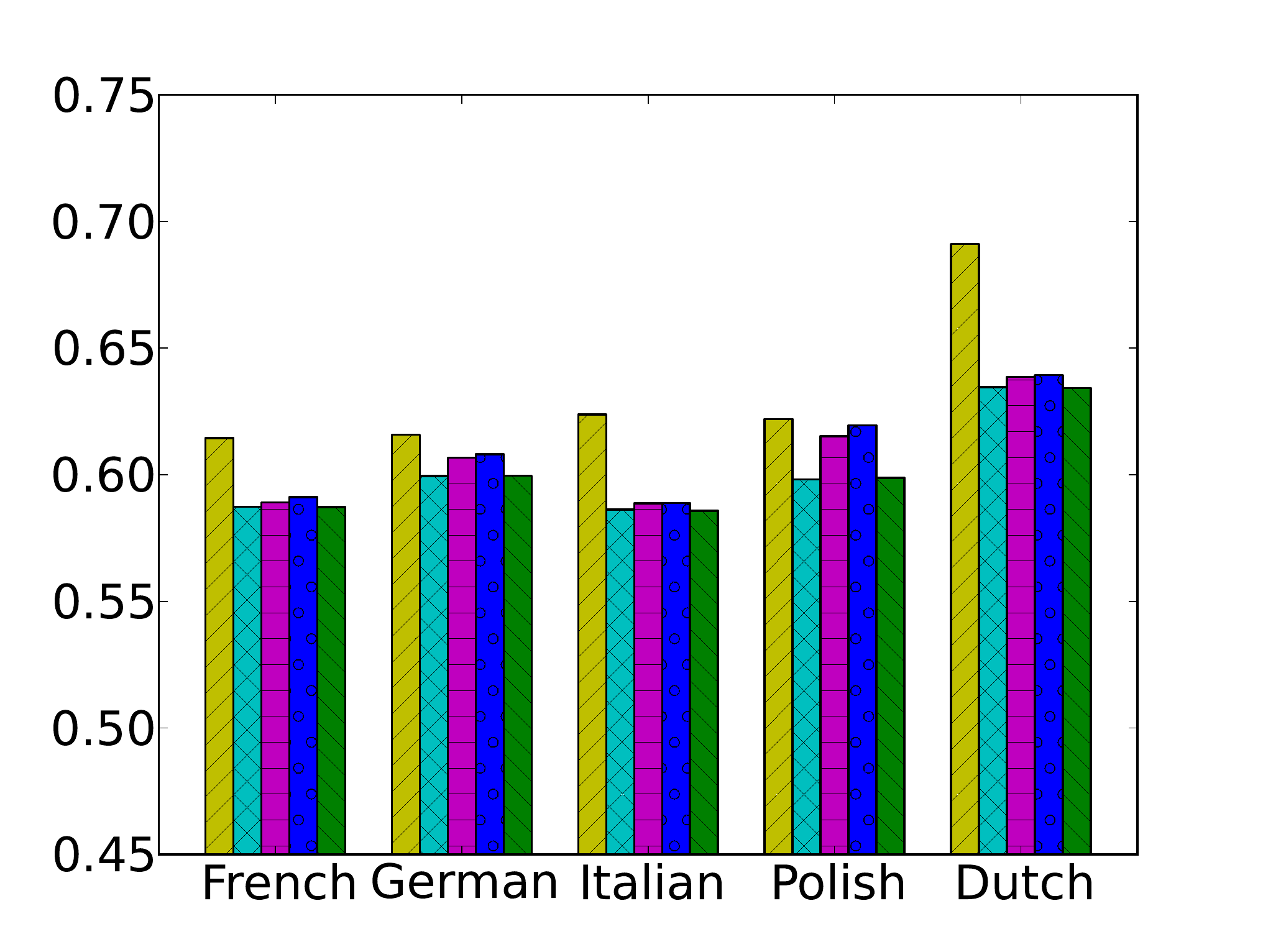, scale=0.25}
\label{fig:precUnlink1}
}
\subfigure[Complement network]{
\epsfig{file=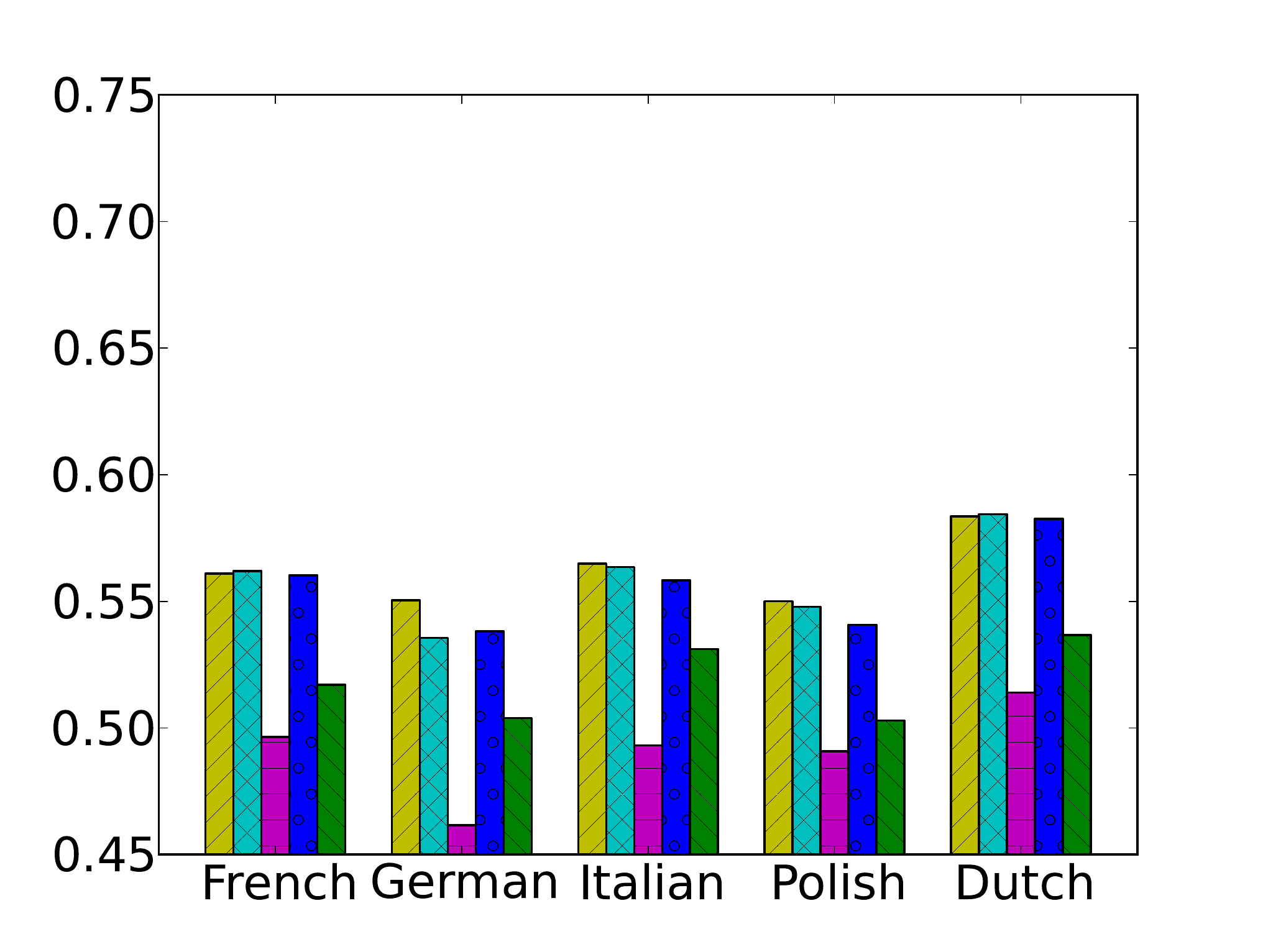, scale=0.25}
\label{fig:precUnlink2}
}
\subfigure{ 
\epsfig{file=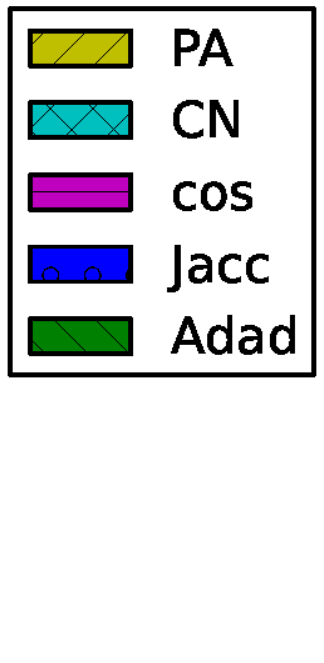, scale=0.35}
}
\caption{Precisions of (a) \textit{complement score} model and (b) \textit{complement network} model. For each method the most successful degree combination is shown.}
\end{figure*}
Figure \ref{fig:precUnlink1} and Figure \ref{fig:precUnlink2} show the best average precisions over all degree combinations of each method for the \textit{complement score} model and the \textit{complement network} model.

The \textit{complement score} model performs significantly better than random, all methods have a precision above 0.5.
Preferential attachment is the top-performing method, superior over the four remaining methods on all five datasets.
This means that the likelihood of an edge to be removed is bigger if the two adjacent nodes have a low degree. 
Up to 69.7\% of all edges from the test set where correctly classified as to remove.
Jaccard and cosine as well as common neighbor and Adamic--Adar perform very similar to each other with precisions above the random baseline, too.

The \textit{complement link prediction} model's precision, shown in Figure \ref{fig:precUnlink2}, has lower precisions than the preceding approach. However, all methods, except cosine, out-perform the random baseline. PA, CN and Jaccard predict link removals with the highest precision, which leads to up to 58.4\% of correct predictions for the test set.
In comparison, the \textit{complement score} approach does a better job in predicting link decay.
\paragraph{Effect of degree combinations}
Computing link decay scores for all edges $(i,j)$ of the test set, we have tested four different degree combinations (cf.\ Table \ref{table:degrees}) of node $i$'s and node $j$'s in respectively out-degree.
\begin{figure*}[hbt!]
\centering
\subfigure[Complement score model]{
\epsfig{file=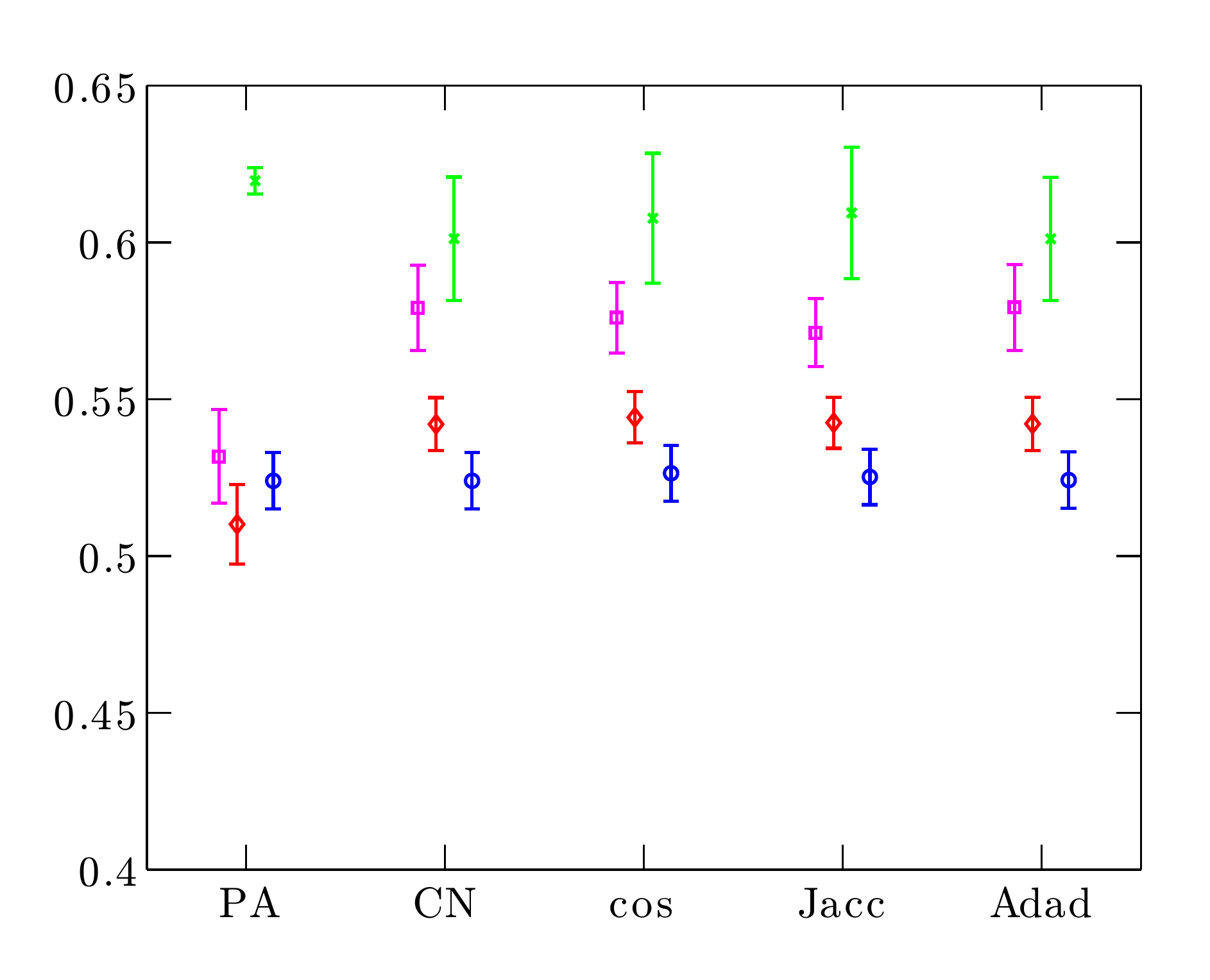, scale=0.3}
\label{fig:degreeCombi1}
}
\subfigure[Complement network model]{
\epsfig{file=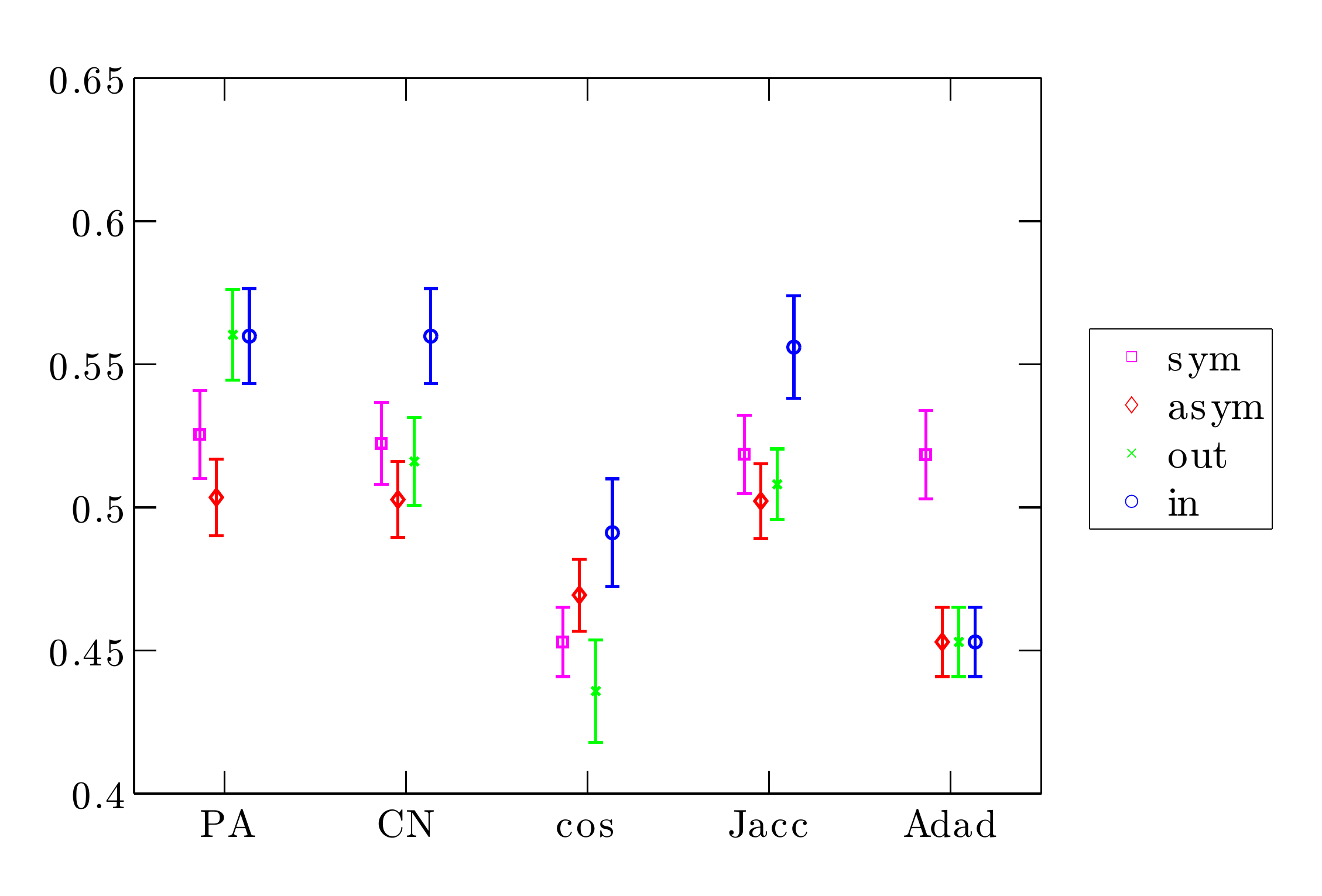, scale=0.3}
\label{fig:degreeCombi2}
}
\caption{Error plot of the four different degree combinations for (a) \textit{complement score} model and (b) \textit{complement network} model. The error is computed by the standard deviation of precision values of the five Wikipedia datasets.}
\label{fig:degreeCombis}
\end{figure*}

In Figure \ref{fig:degreeCombis} we compare the decay prediction
precisions of these four degree combinations across all methods for the
\textit{complement score} approach and the \textit{complement network}
approach. The error bars indicate the standard deviation across the five datasets. 

Varying the types of degrees leads to a drastic deviation within each prediction method. 
For the \textit{inverse link prediction} method, precision values go
from slightly above the random baseline --~when using in-degrees~-- up to 0.6 or more when out-degrees are considered.
The precision values in Figure \ref{fig:degreeCombi1} are staggered: out-degrees perform best, followed by degrees, out-degree/in-degree and in-degrees.
By construction, the \textit{complement network} method thus performs best when considering node in-degrees.
The deviation of precision is not as big as for the inverse method and the ranking of degree combinations is more mixed.
\subsection{Interpretation and Discussion}\label{subsec:discussion}
Our evaluations show that structural analysis makes a meaningful contribution for the prediction of link decays.
Using link prediction methods we have outperformed a random predictor. 
In our evaluations the \textit{complement score} approach combined with preferential attachment performs best. 
Thus, an edge between nodes with a small degree more likely disappears. 
Reasons for this could be because these articles are still evolving, thus their network structure changes because they are not `settled' yet, or, that wrong connections were made caused by the lack of understanding of the article content.
Using only out-going node characteristics, such as a node's out-degree and out-going neighborhood achieved the best precisions for the \textit{complement score} approach. This could be interpreted as some kind of `you are who you link to' rule. Two articles are more similar if they link to the same articles. For link removal this means, that two articles linking to very few common pages  should be dissimilar and thus they should not be connected by a link.

To ascertain whether link decay prediction is of the same difficulty as link prediction, we have also computed link prediction precisions for the five Wikipedia datasets. Actually, link predictions with the same methods are more accurate, precisions around the 0.85 mark were achieved. Thus, the problem of predicting link removals seems to be more difficult than link prediction. 
\paragraph{Weak ties}
The best-performing decay prediction method does not use any community characteristics, such as the number of common neighbors or the union of neighbors.
In the beginning, we have hypothesized that two nodes should not be connected anymore if they have a low degree or if they have a higher degree and have only very few neighbors in common. The first hypothesis is somehow verified by the good precision value of preferential attachment. On the other hand, few neighbors seem not to be a good indicator for link removal. Thus, the network data must contain not only few adjacent nodes with little common neighbors that stay connected.
These links are weak links following Granovetter \cite{granovetter}, that introduce shortcuts into the network which lead to the small-world phenomenon. Considering solely the structure, one cannot distinguish between links that should be removed and links that operate as weak ties.
\section{Conclusions}
In this paper we investigated the problem of predicting the decay in networks such as the Web. We proposed two approaches that utilize link prediction methods and rely on inverted problem descriptions of the link prediction problem. While our first approach simply complements the prediction scores of a link prediction method our second approach applies link prediction to the complement network. Our evaluation showed that, in general, the first approach outperforms the second. However, despite the fact that our evaluation showed that our approaches both outperform the random baseline we discovered that the problem of predicting removal of links is generally harder than the problem of link prediction. 
This observation also justifies the need for further research on the problem of link decay.

To our knowledge this work is the first that investigates the problem of link decay using structural methods in a general manner and the first that investigates the duality of predicting the creation of links and the removal of links.
Ongoing work consists of an even broader evaluation that also takes other network types, such as social networks, into account, and applying further link prediction criteria, such as paths of lengths three and four. 

As future work, a better understanding of further factors beyond existing models and approaches is required.
To especially overcome the issue of distinguishing `weak ties' from `wrong links', the content of the network nodes may also have to be taken into account. 
\jp{Unschoen. Da muss noch mehr hin. Aber nicht zu viel.}

\mt{adjust this section appropriately after the evaluation section has been finished}

\section*{Acknowledgments}
The research leading to these results has received funding from the European Community's 
Seventh Frame Programme under grant agreement n\textsuperscript{o}~257859, 
ROBUST.


\bibliographystyle{abbrv}
\bibliography{bib,ref,kunegis}
\end{document}